\documentclass[prl,amsmath,amssymb,twocolumn,superscriptaddress,showpacs]{revtex4}

\usepackage{amsmath,amssymb}
\usepackage[usenames]{color}
\usepackage{amssymb}
\usepackage{grffile}
\usepackage[pdftex]{graphicx}
\usepackage{amsmath, amstext, amssymb, amsfonts, amsxtra}
\usepackage{textcomp}
\usepackage{xspace} 

\newcommand{\no}{\hat{n}}

\newcommand{\rhop}{\hat{\rho}}             
\newcommand{\rhoss}{\hat{\rho}_{ss}} 
\newcommand{\aop}{\hat{a}} 
\newcommand{\aod}{\hat{a}^{\dagger}}    
\newcommand{\Ho}{\hat{H}}   
\newcommand{\aver}[1]{\langle #1\rangle}

\newcommand{\sutd}{Singapore University of Technology and Design, 8 Somapah Road, 487372 Singapore} 
\newcommand{\majulab}{MajuLab, CNRS-UNS-NUS-NTU International Joint Research Unit, UMI 3654, Singapore} 
\newcommand{\cqt}{Centre for Quantum Technologies, National University Singapore, Singapore 117543, Singapore}
\newcommand{\nus}{Physics Department, National University of Singapore, 2 Science Drive 3 Singapore 117551, Singapore}

\begin{document}

\title{Tuning energy transport using interacting vibrational modes}                     

\author{Chu Guo} 
\affiliation{\sutd}
\author{Manas Mukherjee}
\affiliation{\cqt}
\affiliation{\nus} 
\affiliation{\majulab} 
\author{Dario Poletti}
\affiliation{\sutd}
\affiliation{\majulab}

\begin{abstract} 
We study energy transport in a chain of quantum harmonic and anharmonic oscillators where the anharmonicity is induced by interaction between local vibrational states of the chain. Using adiabatic elimination and numerical simulations with matrix product states, we show how strong interactions significantly slow down the relaxation dynamics (with the emergence of a new time scale) and can alter the properties of the steady state. We also show that steady state properties are completely different depending on the order in which the limits of infinite time and infinite interaction are taken.   
\end{abstract}


\maketitle

Miniaturization of devices is bound to reach the limits set by quantum mechanics in terms of both device output and heat/energy management. Therefore characterizing and controlling energy transport in nanostructures is critical for the future engineering and design of quantum materials and devices. Fundamental studies in both theory and experiments are required. A promising approach for these studies is to analyze prototypical models which can simulate condensed matter systems and can be implemented experimentally in a clean, controlled and highly tunable manner \cite{ReviewOnIons,ReviewOnIons2, BlochZwerger2008}. As recent theoretical proposals \cite{BermudezPlenio2013} and experimental achievements \cite{HazeUrabe2012, RammHaffner2013, SenkoMonroe2014, AnKim2015} have shown, cold ions are a great candidate for studying energy flow. In fact the ions form a regular artificial crystalline structure and each ion can vibrate around its equilibrium position thus affecting the vibration of the neighboring ions via Coulomb interaction. The use of site-resolved addressing allows control of the probability distribution of the different local vibrational modes. Moreover, in these set-ups it is possible to measure, both locally and globally, the excitation of the vibrational modes. 

Energy and particle transport in quantum boundary driven one-dimensional chains have been studied for spins, fermions and non-interacting bosons. Some notable effects include negative differential resistance \cite{Prosen2007,BenentiRossini2009,BenentiZnidaric2009}, emergence of long range order \cite{ProsenZnidaric2010}, many-body resonances related to extended states in the cavity array \cite{BiellaFazio2015}, existence of more steady states with quantum-number conserving baths \cite{BucaProsen2012}, study of mesoscopic \cite{AjisakaProsen2012} or time-dependent baths \cite{ZnidaricProsen2011, ProsenIlievski2011} and exact solutions in systems whose bulk is integrable \cite{ProsenZunkovic2010, Prosen2011, Prosen2011b, Prosen2014, PopkovProsen2015, Prosen2015}.  

In this work we show the influence of strong interactions on energy flow, relaxation dynamics and on the steady state for a bosonic chain (as it can be realized in cold ions experiments \cite{BermudezPlenio2013, PorrasCirac2004}). We first analyze the scenario in which the site-specific tailored dissipation produced by simultaneous application of red and blue sideband lasers acts only at one extremity of the chain. Henceforth we refer to it as `tailored bath' to differentiate it from a thermal dissipative bath. We show that in this first scenario the steady state is independent of the strength of the interaction applied to other sites, in contrast to the case in which the system is connected to a realistic thermal bath. The scenario with one single site connected to the tailored bath is also ideal to show how strong interactions significantly slow down the relaxation dynamics generating a new time scale which increases with increasing interaction strength. We then proceed to study the boundary driven case (i.e. the sites at the boundaries are driven by two tailored baths). In this case we show how interactions break ballistic transport and can be used to tune (but not to eliminate) the local bosonic occupation as well as the particle and energy currents in the chain. We stress that, in the strongly interacting regime, the emergent non-equilibrium physics is very different from that of hard core bosons (i.e. bosonic particles with infinite repulsive interaction).     


\begin{figure}
\includegraphics[width=\columnwidth]{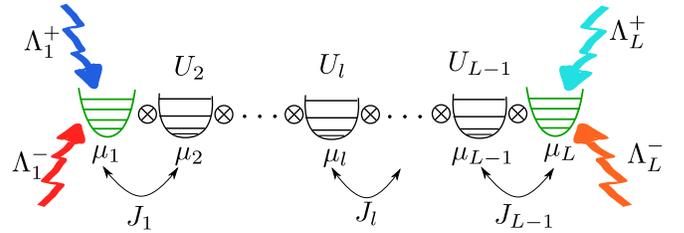}
\caption{(Color online) Depiction of the system under study. A chain of oscillators in which the ones at the extremities are local harmonic oscillators while those in the middle are anharmonic. The local oscillators are described by bosonic modes characterized by a local potential $\mu_l$ and local interaction strength $U_l$. Vibrations in each oscillator can tunnel to neighboring sites via the tunneling $J_l$. The harmonic oscillators are in contact with an environment through couplings $\Lambda^+_{\sigma}$ and $\Lambda^-_{\sigma}$. } \label{fig:summary} 
\end{figure}

The system we analyze is a chain of $L$ cold ions forming a linear lattice where vibrational quanta are transported along the chain. Since the vibrational quanta of the local ion oscillator are bosons, we will refer to them as vibron \cite{BermudezPlenio2013}. The dynamics can be described by 
\begin{equation} 
\frac{d}{dt}\rhop = - \frac i {\hbar} \left[\Ho,\rhop\right] + \mathcal{D}(\rhop) \label{eq:lind}     
\end{equation}
where $\rhop$ is the density matrix of the system, $\Ho$ is the Hamiltonian and $\mathcal{D}$ is the dissipative part of the evolution \cite{BermudezPlenio2013}. 
The Hamiltonian $\Ho$ is given by \cite{PorrasCirac2004, DuttaSengupta2013, DuttaSengupta2015} 
\begin{equation} 
\Ho =  \sum_{l=1}^{L-1} \left( J_l\aod_l\aop_{l+1} + {\rm h.c.} \right) + \sum_{l=1}^{L} \left[ \frac{U_l}{2}\no_l(\no_l-1) + \mu_l \no_l \right]            
\end{equation}  	
where $J_l$ is the tunneling amplitude from site $l$ to site $l\pm 1$, $U_l$ is the local interaction amplitude (due to induced anharmonicity from external lasers \cite{PorrasCirac2004}) and $\mu_l$ is the local potential (while $\mu_l$ can be locally controlled, we keep it constant throughout this work $\mu_l=\mu$). In the remainder of the article we will take $J_l=J$ constant throughout the chain. The operators $\aop_l$ and $\aod_l$ respectively destroy or create a vibron at site $l$.    
The tailored baths act locally on the harmonic oscillators injecting and extracting vibrons from two extremities: 
\begin{eqnarray} 
\mathcal{D}(\rhop) =  \sum_{l=1,L}&& \left[ \frac{\Lambda_{l}^-}{2} \left(2\aop_l \rhop \aod_l - \{\aod_l\aop_l,\rhop\} \right) \right. \nonumber \\ 
&&+ \left. \frac{\Lambda_{l}^+}{2} \left(2\aod_l \rhop \aop_l - \{\aop_l\aod_l,\rhop\} \right) \right]  
\end{eqnarray}  
Here $\Lambda_l^+$ ($\Lambda_l^-$) is the heating (cooling) rate of the dissipation. 
The system that we study is schematically represented in Fig.\ref{fig:summary} which shows, for example, two harmonic oscillators coupled to the tailored environment at the boundaries and a bulk of anharmonic oscillators which are generated by local modification of the potential. The size of reduced Hilbert space that we consider scales as $S=d^{2L}$ where $d$ is the number of local vibrational modes considered. In our simulations $d\ge 10$ which for $L=4$ corresponds to $S\ge 10^8$. We will thus use analytical and numerical methods based on adiabatic elimination \cite{GarciaRipollCirac2009, PolettiKollath2012, PolettiKollath2013, SciollaKollath2015} and the time-dependent Matrix Product States (t-MPS) algorithm \cite{Vidal2003, DaleySchollwock2004, White2004, Schollwock2011}.   


\begin{figure}
\includegraphics[width=\columnwidth]{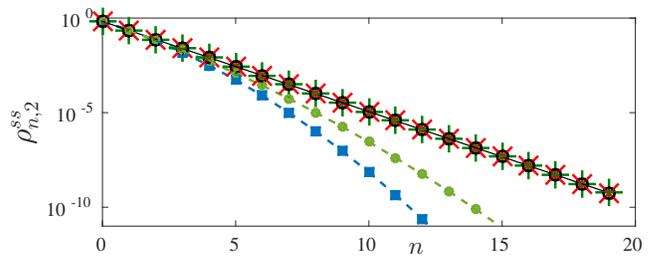}
\caption{(Color online) Diagonal elements of the density matrix of the second site at steady state $\rho^{ss}_{n,2}$. The continuous black line represents the theoretical prediction; numerical results are represented by the symbols: black circle $\circ$ for $U=0$, red $\times$ for $U=3J$ and green $+$ for $U=6J$. The dissipation on the left sites are $\hbar\Lambda_1^+ = 3 J$, $\hbar\Lambda_1^- = 6 J$. The dashed lines show the distribution for a local thermal distribution for $U=3J$ (green circles) and $U=6J$ (blue squares).} \label{fig:twoions} 
\end{figure}

{\it A single bath}: The steady state, $\rhoss$, for a chain in which only one site at an extremity is connected to the tailored environment is found to be a product state $\rhoss=\bigotimes_l \left(\sum_n \rho^{ss}_{n,l} |n,l\rangle \langle n,l|\right)$ where $|n,l\rangle$ identifies the $n$-th vibrational state on site $l$ and $\rho^{ss}_{n,l}$ the probability of it being occupied \cite{supplementary, nosteadystates}. 
The value of $\rho^{ss}_{n,l}$ is independent of the site $l$, the local interaction $U_l$ and the local potential $\mu_l$. It only depends on the amplitudes of the tailored dissipative coupling and it is given by 
\begin{equation} 
\rho^{ss}_{n,l}= \sum_{n=0}^{\infty}\frac{(\Lambda^{+}_1/\Lambda^{-}_1)^n}{(1-\Lambda^{+}_1/\Lambda^{-}_1)^{-1}}
\end{equation}
This is clearly shown in Fig.\ref{fig:twoions} for the illustrative case of a two site chain in which the first site (from the left) is coupled to the tailored dissipation while the second site can have different values of the interaction. The figure depicts the steady state occupation of the vibrational modes \cite{approx} on the right site for different values of the interaction $U_2=0,\;3,\;6$ respectively by the black circles $\circ$, red $\times$ and green $+$. We note that the numerical results agree exactly with the theoretical expectation without any fitting parameter. The dashed lines with filled circles ($U_2=3$) and squares ($U_2=6$) represent instead the values of $\rho^{ss}_{n,2}$ if the final state was a local thermal state with the same ``temperature'' as the first site. In this case the occupation of the local vibrational levels depends on the energy gap between the different levels and thus on the strength of the local interaction (anharmonicity). Hence, the dissipation studied here is not a thermal bath but a tailored dissipative process which tries to force a certain probability distribution of occupations in the energy levels of the density matrix independent of their energy. 

At this point it is possible to highlight a feature of this tailored dissipative system which will also be very important in the following. Suppose that the tailored dissipation would be driving the system towards a product of local thermal states with the same temperature. In that case, when the repulsive interaction becomes significantly large, the anharmonic oscillators could be very well described by just the bottom two levels (also known as the hard core bosons limit) because the population of any other level would be significantly suppressed. However the actual tailored dissipation used imposes a particular local distribution independent of the vibrational energy levels. This means that very high energy levels would still be highly occupied and the hard core bosons picture cannot be used. In other words, the limits $\lim_{t\rightarrow\infty}$ and $\lim_{U_l\rightarrow\infty}$ do not commute. The fact that high  energetic levels are occupied also strongly affects the relaxation dynamics of the system as we analyze in the following. 

\begin{figure}
\includegraphics[width=\columnwidth]{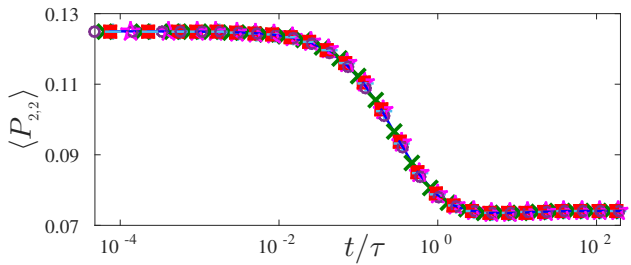}
\caption{(Color online) Probability of occupying level $2$ of the oscillator on site $2$, $\langle P_{2,2}\rangle$ versus time. The parameters used are: $U=30 J$ ,$\hbar\Lambda_1^+ = 3J$, $\hbar\Lambda_1^- = 9J$ for the pink stars; $U=40J,\hbar\Lambda_1^+ = 3J, \hbar\Lambda_1^- = 9J$ for the red filled squares; $U=50J$, $\hbar\Lambda_1^+ = 3J$, $\hbar\Lambda_1^- = 9J$ for the purple circles; $U=50J$, $\hbar\Lambda_1^+ = 4J$, $\hbar\Lambda_1^- = 12J$ for the green $\times$; $U=50J$, $\hbar\Lambda_1^+ = 5J$, $\hbar\Lambda_1^- = 15J$ for the light-blue $+$.} \label{fig:timescale} 
\end{figure}

{\it Two sites dynamics}: As initial condition we consider a chain prepared in a tensor product of local thermal states with $U_l=0$ $\;\forall \; l$ and with constant average density $\aver{\no_l}=\aver{\no_m}=1\;\forall\; l,m$. Here $\aver{\hat{O}}={\rm tr}(\rhop \hat{O})$ is the trace over the density matrix $\rhop$ of the operator $\hat{O}$. Subsequently, at time $t=0$ the left site is coupled to a tailored dissipative process that would impose a local average density $\aver{\no_1 (t=\infty)}=0.5$ lower than the initial density. At the same time, the interaction on site $2$, $U_2$, is also abruptly switched on. In the regime $J \ll \Lambda_{l}^{+}, \Lambda_{l}^{-}, l = 1, L$, we can adiabatically eliminate the dynamics of the site in contact with tailored environment, and only consider the time evolution of the rest of the system (which, in this case, is the second ion) \cite{supplementary, GarciaRipollCirac2009, PolettiKollath2012, PolettiKollath2013, SciollaKollath2015}. The equation of motion for the diagonal elements of the reduced density matrix on the second site, $\rho_{n,2}$, is
\begin{eqnarray}
\frac{d \rho_{n,2}(t)}{d t} &=& -\left(\chi^{1,-}_{n,n-1}+\chi^{1,+}_{n+1,n}\right) \rho_{n,2}(t) \label{eq:2sitesad}   \\ 
\nonumber &+&  \chi^{1,-}_{n+1,n} \; \rho_{n+1,2}(t) +  \chi^{1,+}_{n,n-1}  \; \rho_{n-1,2}(t) 
\end{eqnarray}
where $\chi^{l,\sigma}_{m,n}=2J^2\Lambda^{\sigma}_{l}m/\left[U^2_2n^2+\hbar^2\left(\Lambda_l^+ - \Lambda_l^-\right)^2\right]$. 
From Eq.(\ref{eq:2sitesad}), which can easily be solved numerically, we learn of the presence of a new time scale $\tau=U_2^2/\left[J^2 \left(\Lambda^-_l - \Lambda^+_l\right)\right]$ (when $U_2 \gg \hbar(\Lambda^-_l-\Lambda^+_l)$). This result is clearly confirmed by Fig.\ref{fig:timescale} in which we plot the probability of having two vibrons at the second site given by $\aver{P_{2,2}}$ for different values of both the interaction $U_l$ and coupling to the environment $\Lambda^{\sigma}_l$. All the curves collapse on top of each other once the time is rescaled with the new, emerging, time scale $\tau$.  

\begin{figure}
\includegraphics[width=\columnwidth]{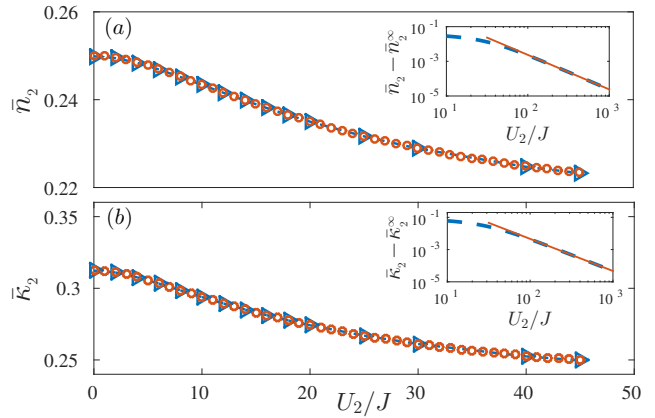}
\caption{(Color online) (a) $\bar{n}_2$  and (b) local fluctuations $\bar{\kappa}_2$ as a function of $U_2$: the blue triangles are results from the numerical t-MPS method while the red circles are analytic predictions by adiabatic elimination. The dissipation coefficients used on the first and third sites are $\hbar\Lambda_1^+ = 5 J, \hbar\Lambda_1^- = 21.66 J$ and $\hbar\Lambda_3^+ = 5J$ and $\hbar\Lambda_3^- = 55J$. The insets represent the values of the density or fluctuations minus their asymptotic values for $U\rightarrow\infty$, $\bar{n}_2^{\infty}$ and $\bar{\kappa}_2^{\infty}$. The blue dashed lines represent the numerical results while the red continuous lines are power-law fit with the exponent $-2$.} \label{fig:3sites} 
\end{figure}

{\it Boundary driven case}: Finally, we consider the case in which the chain is coupled to two tailored environments at its extremities. Each boundary is coupled differently, such that vibrons and energy currents are imposed. We show that, unlike the scenario with only one site coupled to the tailored environment, the local interaction strongly affects the profile of the vibron occupation, the particle current and the energy current in the system. We observe first the case of a chain composed of only $3$ sites, hence the interaction is only applied to the central site, $U_2\ne 0$ while $U_1=U_3=0$. In Fig.\ref{fig:3sites}(a) and (b) respectively, we show how the local density $\bar{n}_2=\langle\hat{n}_2\rangle_{ss}$ and local fluctuations $\bar{\kappa}_2=\langle\hat{n}^2_2\rangle_{ss}-\langle\hat{n}_2\rangle^2_{ss}$ of the steady state ($ss$) change with the interaction strength $U_2$. These plots show a remarkable agreement between the exact simulation of the $3$ sites (blue triangles) and the adiabatic elimination approach (red circles). 
%
For $3$ sites, the equations stemming from the adiabatic elimination are given by 
\begin{eqnarray} 
\frac{d \rho_{n,2}(t)}{d t} &=& -\left(\sum_{\substack{{l=1,3}\\ {\sigma=\pm 1}}}\!\!\!\!\chi^{l,\sigma}_{n,n-1}\! + \chi^{l,\sigma}_{n+1,n} \right)  \rho_{n,2}(t)  \nonumber \\ 
&&+\sum_{{l=1,3}}\chi^{l,-}_{n+1,n}  \rho_{n+1,2}(t)  \nonumber \\
&&+\sum_{{l=1,3}}\chi^{l,+}_{n,n-1}  \rho_{n-1,2}(t)
\label{eq:ae3}
\end{eqnarray} 
The particular structure of these equations make it easy to compute the steady state \citep{supplementary} 
\begin{equation}
\rho^{ss}_{n,2}= \rho_{0,2}^{ss} \prod_{k=0}^{n-1}\frac{\sum_{l=1,3} \chi^{l,+}_{k+1,k}}{\sum_{l=1,3} \chi^{l,-}_{k+1,k}} 
\end{equation}
where $\rho_{0,2}^{ss}$ sets the normalization. This expression agrees with our numerical simulations performed with t-MPS \cite{implementation}. 
%
The insets of Fig.\ref{fig:3sites}, computed using Eq.(\ref{eq:ae3}), show that the values of $\bar{n}_2$ and $\bar{\kappa}_2$ approach their respective asymptotic value for $U_2\rightarrow\infty$, $\bar{n}_2^{\infty}$ and $\bar{\kappa}_2^{\infty}$, as a power law with an exponent which, as predicted from Eq.(\ref{eq:ae3}), is equal to $-2$. The asymptotic values are $\bar{n}_2^{\infty}=\beta_1/[(1-\beta_2)(1+\beta_1-\beta_2)]\approx 0.215$ and $\bar{\kappa}^{\infty}_2 = \bar{n}_2^{\infty} \left(1+\beta_2\right)/\left(1-\beta_2\right) - \left(\bar{n}_2^{\infty}\right)^2\approx 0.233$, where $\beta_1 = \left[\frac{\Lambda_1^+}{(\Lambda_{1}^{-}-\Lambda_{1}^{+})^2}+\frac{\Lambda_3^+}{(\Lambda_{3}^{-}-\Lambda_{3}^{+})^2}\right]/\left[\frac{\Lambda_1^-}{(\Lambda_{1}^{-}-\Lambda_{1}^{+})^2}+\frac{\Lambda_3^-}{(\Lambda_{3}^{-}-\Lambda_{3}^{+})^2}\right]$, and $\beta_2 = \left(\Lambda_1^++\Lambda_3^+\right)/\left(\Lambda_1^-+\Lambda_3^-\right)$. Note that a hardcore bosons approach would predict $\bar{n}_2^{_{\rm HC}}\approx 0.167$ and  $\bar{\kappa}_2^{_{\rm HC}}=0$.                     

%
\begin{figure}
\includegraphics[width=\columnwidth]{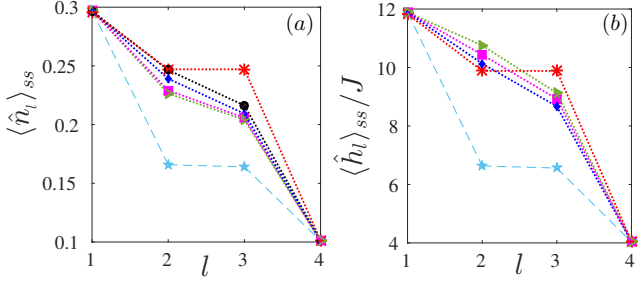}
\caption{(Color online) Local density $\langle\hat{n}_l\rangle_{ss}$ (a) and local energy $\langle\hat{h}_l\rangle_{ss}$ (b) versus the site number for a chain of $4$ sites. The values of the interaction strength $U_2=U_3$ are $U_2=0$ red stars, $U_2=5 J$ black circles, $U_2=10 J$ blue diamonds, $U_2=25 J$ purple squares and $U_2=35 J$ brown triangles. The dashed line represent the results from an hardcore bosons study. Other parameters are $\hbar\Lambda^+_1=J$, $\hbar\Lambda^-_1=4.33J$, $\hbar\Lambda^+_3=J$ and $\hbar\Lambda^-_3=11J$.} \label{fig:dens_4_5} 
\end{figure}
Using MPS we further investigate a chain of $4$ ion sites which would allow us to study how the local vibron density $\langle\hat{n}_l\rangle_{ss}$ and local energy $\langle\hat{h}_l\rangle_{ss}$ are modified from the ballistic transport (typical of $U_l=0$) to a new regime induced by the local interaction. The local energy is defined as $\hat{h}_l=J/2\left(\hat{a}_l^+\hat{a}_{l-1}+\hat{a}_l^+\hat{a}_{l+1}+h.c.\right)+(U_l/2)\hat{n}_l(\hat{n}_l-1)+\mu\hat{n}_l$. In Fig.\ref{fig:dens_4_5} we show that the interaction breaks the ballistic transport, the local occupation of the levels decreases and moreover it is not constant along the chain. The local energy, at large $U_l$ is dominated by the interaction and can be larger than the non-interacting case $U_l=0$. The dashed lines in Fig.\ref{fig:dens_4_5} show the result of a hardcore bosons analysis in which the limit $U\rightarrow \infty$ is taken before $t\rightarrow\infty$. 

\begin{figure}
\includegraphics[width=\columnwidth]{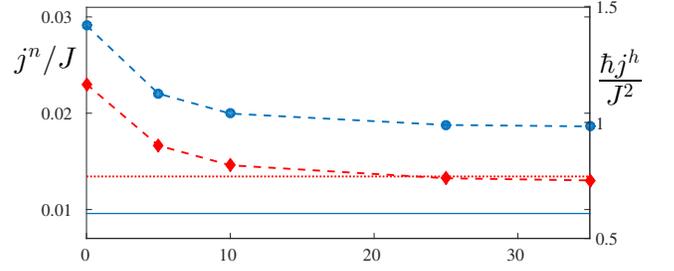}
\caption{(Color online) Particle current $j^n$ (blue circles) and energy current $j^{h}$ (red diamonds) versus the site number for a chain of $4$ sites. The red dotted and the continuous blue lines represent the hardcore bosons limit respectively for the energy and the particle currents. Other parameters are $\hbar\Lambda^+_1=J$, $\hbar\Lambda^-_1=4.33 J$, $\hbar\Lambda^+_3=J$, $\hbar\Lambda^-_3=11J$ and number of sites $L=4$.} \label{fig:current} 
\end{figure}

We then turn to study how the interaction changes the current flow in the system. We analyze both particle current $j^n$ and energy current $j^h$. Their expression is derived using the continuity equation 
\begin{eqnarray}
\hbar \partial_t \hat{O}_l = j^O_{l-1,l} - j^O_{l,l+1} +S_l, 
\end{eqnarray} 
where $j^O_{l-1,l}$ is incoming current, $j^O_{l,l+1}$ outgoing (both related to the Hamiltonian dynamics) and $S_l$ is a source term (due to dissipation).  
This results in 
\begin{align} 
j^n_{l,k}&=i\frac{J}{\hbar}\langle\hat{a}^{\dagger}_k\hat{a}_{l} - \hat{a}^{\dagger}_{l}\hat{a}_k\rangle \\ 
j^h_{l-1,l}&=\frac{iJ}{2\hbar}\langle (U_{l-1}\hat{n}_{l-1}+U_{l}\hat{n}_l) \hat{b}_{l-1}\hat{b}_{l}^{\dagger}  \nonumber \\ 
 &- \hat{b}_{l}\hat{b}_{l-1}^{\dagger} (U_{l-1}\hat{n}_{l-1}+U_{l}\hat{n}_l) \rangle  \nonumber \\ 
 &- \frac{J}{2\hbar} \hat{j}_{l-2,l}+\left(\frac{\mu}{\hbar}-\frac{U_{l}}{2\hbar}\right)j_{l-1,l}
\end{align}
In steady state the current is independent of the site and we can use lighter notations $j^n$ and $j^h$. Fig.\ref{fig:current} shows how both currents are lowered by the interaction towards an asymptotic limit different from hardcore bosons predictions.

{\it Experimental realization}: The system studied here can be implemented using a chain of cold ions in a linear trap~\cite{PorrasCirac2004}. Though the results obtained here are independent of the specificities of the ion involved, the actual values of $J$ and $U_l$ depend on the mass and internal energy levels of the ions involved. For Barium ions, $^{138}$Ba$^+$, it is possible to achieve an on-site interaction strength $U_l\approx 5J$ with reasonable laser power~\cite{DuttaSengupta2013}. Although going beyond $8J$ may not be feasible using a standing wave configuration as proposed by Porras~\cite{PorrasCirac2004},  the effects of interaction on  transport is quite pronounced even at lower interaction strength (see Figs. \ref{fig:dens_4_5} and \ref{fig:current}). The tailored bath has been implemented in an ion trap setup~\cite{RammHaffner2013} while hopping of transverse vibron has been implemented by Haze et al.~\cite{HazeUrabe2012}. Addressing individual sites for a chain of $4$ ions with inter-ionic distances as large as tens of microns does not seem to pose any experimental challenge, however for a large number of individually addressed ions, laser beams with finite focussing may affect the neighbouring ions. Therefore we believe, with present day state-of-the-art techniques it is possible to realize the system proposed here as a proof-of-principle for a chain below ten ions.

{\it Conclusions}: We have shown that anharmonicity induced interactions break the ballistic transport of particles and energy in the non-interacting regime and can be used to regulate the time scales and the steady state of energy transport in linear chains. We have developed an analytical understanding of these phenomena thanks to adiabatic elimination and shown that the properties of the system in the strong interaction limit are very different from a hardcore bosons approach. This work is an important step towards the understanding of energy and heat flow in interacting bosonic systems.      

We are thankful to U. Bissbort, S. Denisov, S. Flach, M. Fleischhaur, Z. Gang, N. Li, D. Rossini and C. Teo for insightful discussions. This project is funded by SUTD Start-up grant EPD2012-045.



\newpage 
\section{Supplementary Material}   

\subsection{Steady state for a chain with one boundary site under dissipation} 
In this section we aim to show that Eq.(4) of the main article is indeed a steady state of the system. 
First of all we note that $\mathcal{D}(\hat{\rho}_{ss})=0$. In fact 
\begin{align}
\mathcal{D}(\hat{\rho}_{ss})&=\sum_{1}^{\infty} \Lambda_1^-\frac{(\Lambda_1^+/\Lambda_1^-)^n}{1-\Lambda_1^+/\Lambda_1^-} n|n-1\rangle\langle n-1| \nonumber\\ 
&- \sum_{0}^{\infty} \Lambda_1^-\frac{(\Lambda_1^+/\Lambda_1^-)^n}{1-\Lambda_1^+/\Lambda_1^-}n |n\rangle\langle n| \nonumber\\ 
&+ \sum_{0}^{\infty} \Lambda_1^+\frac{(\Lambda_1^+/\Lambda_1^-)^n}{1-\Lambda_1^+/\Lambda_1^-}(n+1)|n+1\rangle\langle n+1| \nonumber\\ 
&- \sum_{0}^{\infty} \Lambda_1^+\frac{(\Lambda_1^+/\Lambda_1^-)^n}{1-\Lambda_1^+/\Lambda_1^-}(n+1)|n\rangle\langle n| \tag{1S} \label{eq:Drho}
\end{align}  
where the first line in Eq.(\ref{eq:Drho}) cancels the fourth and second is equal and opposite to the third.    
It is also possible to show that $[\hat{H}, \hat{\rho}_{ss}]=0$. The interacting and the potential part of the Hamiltonian are readily verified. As for the kinetic part of the Hamiltonian, it is sufficient to show that  $[\hat{a}^{\dagger}_l\hat{a}_p, \hat{\rho}_{ss}]=0$. Indeed you have   
\begin{align}
\nonumber &\hat{a}^{\dagger}_l\hat{a}_p 
\underbrace{\left[\sum_{m,n=0}^{\infty}|m \rangle_l\langle m|\bigotimes |n \rangle_p\langle n| \frac{(\Lambda_1^+/\Lambda_1^-)^{m+n}}{(1-\Lambda_1^+/\Lambda_1^-)^{-2}}\right]}_{{\normalfont \hat{\rho}_{ss}}}\\ 
&=\left[\sum_{m=0}^{\infty}\sum_{n=1}^{\infty}\sqrt{n(m+1)}|m+1 \rangle_l\langle m| \right.  \nonumber \\
& \left. \bigotimes |n-1 \rangle_p\langle n| \frac{(\Lambda_1^+/\Lambda_1^-)^{m+n}}{(1-\Lambda_1^+/\Lambda_1^-)^{-2}}\right]\nonumber \\ 
&=\Lambda_1^+/\Lambda_1^-\left[\sum_{m,n=0}^{\infty}\sqrt{(n+1)(m+1)}|m+1 \rangle_l\langle m| \right.  \nonumber \\
& \left. \bigotimes |n \rangle_p\langle n+1| \frac{(\Lambda_1^+/\Lambda_1^-)^{m+n}}{(1-\Lambda_1^+/\Lambda_1^-)^{-2}}\right]\tag{2S} 
\end{align}
and at the same time 
\begin{align}
&\left[\sum_{m,n=0}^{\infty}|m \rangle_l\langle m|\bigotimes |n \rangle_p\langle n| \frac{(\Lambda_1^+/\Lambda_1^-)^{m+n}}{(1-\Lambda_1^+/\Lambda_1^-)^{-2}}\right] \hat{a}^{\dagger}_l\hat{a}_p \nonumber \\ 
&=\left[\sum_{m=1}^{\infty}\sum_{n=0}^{\infty}\sqrt{(n+1)m}|m \rangle_l\langle m-1| \right.  \nonumber \\
& \left. \bigotimes |n \rangle_p\langle n+1| \frac{(\Lambda_1^+/\Lambda_1^-)^{m+n}}{(1-\Lambda_1^+/\Lambda_1^-)^{-2}}\right]\nonumber \\ 
&=\Lambda_1^+/\Lambda_1^-\left[\sum_{m,n=0}^{\infty}\sqrt{(n+1)(m+1)}|m+1 \rangle_l\langle m| \right.  \nonumber \\
& \left. \bigotimes |n \rangle_p\langle n+1| \frac{(\Lambda_1^+/\Lambda_1^-)^{m+n}}{(1-\Lambda_1^+/\Lambda_1^-)^{-2}}\right]\tag{3S}
\end{align}
which is identical to Eq.(2S). 

\subsection{Adiabatic elimination}
Here we show one way to derive Eq.(5) of the main paper. In the regime in which $\Lambda_1^{\sigma}\gg J$ it is possible to write classical diffusion equations for the evolution of the diagonal elements of the single site reduced density matrix $\rho_{n,2}$. We will follow the method described in detail in the supplementary material of \cite{SciollaKollath2015}. 
Using the same notations as \cite{SciollaKollath2015} we divide the Liouvillian evolution in two parts, 
\begin{align} 
\mathcal{L}(\hat{\rho})&=\mathcal{L}_0(\hat{\rho})+\mathcal{V}(\hat{\rho})\tag{4S}\\ 
\mathcal{L}_0(\hat{\rho}) &= -\frac{i}{\hbar} \left[\hat{H}_D,\hat{\rho}\right]+\mathcal{D}(\hat{\rho}) \tag{5S}\\ 
\mathcal{V}(\hat{\rho}) &= -\frac{i}{\hbar} \left[\hat{H}_K,\hat{\rho}\right] \tag{6S}     
\end{align} 
where $\hat{H}_D=U_2/2\sum_{n_2}\hat{n}_2(\hat{n}_2-1)+\mu\sum_{n_1,n_2}\left(\hat{n}_2+\hat{n}_1\right)$ and $\hat{H}_K=-J\left(\hat{b}^{\dagger}_1\hat{b}_2+{\rm h.c.}\right)$. 
Let us take 
\begin{align}
\hat{\rho}^{\Omega_0}=&\left(\sum_{n_2}\rho_{n_2,2}|n_2\rangle\langle n_2|\right)\bigotimes\hat{\rho}^{ss}_1 \nonumber\\ 
=&\left(\sum_{n_2}\rho_{n_2,2}|n_2\rangle\langle n_2|\right)\bigotimes\left(\sum_{n_1}\frac{\alpha^{n_1}}{\beta}|n_1\rangle\langle n_1|\right) \tag{7S}   
\end{align} 
where $\alpha=\Lambda_1^+/\Lambda_1^-$ and $\beta=\left(1-\Lambda_1^+/\Lambda_1^-\right)^{-1}$ and $\rho_{n_2,2}$ has the same meaning as $\rho_{n,2}$ in the main paper. $\hat{\rho}^{\Omega_0}$ lives in the so-called $\Omega_0$, decoherence free, space and it is connected to dissipating spaces $\Omega_{\alpha}$ via the kinetic term of the Hamiltonian which is a small perturbation \cite{OmegaLambda}. 
The time evolution of $\hat{\rho}^{\Omega_0}$ is given by 
\begin{align} 
\frac{d}{dt}\hat{\rho}^{\Omega_0}&=\tilde{\mathcal{L}}\left(\hat{\rho}^{\Omega_0}\right) \tag{8S}\\ 
\tilde{\mathcal{L}}&=\mathcal{L}_0-\sum_{\alpha}\mathcal{V}\left(\mathcal{L}_0\right)^{-1}\mathcal{V} \tag{9S}      
\end{align}   
where, in our case, $\mathcal{L}_0(\hat{\rho}^{\Omega_0})=0$. $\mathcal{V}(\hat{\rho}^{\Omega_0})$ gives $4$ different terms and we choose one of them, $\hat{C}=\left(iJ/\hbar\right)\hat{b}^{\dagger}_1\hat{b}_2\hat{\rho}^{\Omega_0}$, to exemplify the derivation of Eq.(5) of the main paper. This gives  
\begin{align} 
\mathcal{L}_0(\hat{C}) =&  \left(iU_2(n_2-1) - \left(\Lambda_1^--\Lambda^+_1\right)\right)\hat{C} \tag{10S}     
\end{align}   
This results in the diagonal terms of the density matrix to follow the differential equation 
\begin{align} 
\frac{d}{dt}\rho_{n_2,2}&=\langle n_2| {\rm tr}_1 \!\left(\frac{d}{dt}\hat{\rho}^{\Omega_0}\right)|n_2\rangle \nonumber \\ 
&=\langle n_2| {\rm tr}_1 \!\left(\tilde{\mathcal{L}}\hat{\rho}^{\Omega_0}\right)|n_2\rangle \tag{11S}     
\end{align}   
where ${\rm tr}_1$ traces out the first site. After some computation this turns into 
\begin{align} 
\frac{d}{dt}\rho_{n_2,2}=&  \nonumber\\ 
-\sum_{n_1}&\left[\frac{2J^2(n_2+1)n_1(\Lambda_1^--\Lambda_1^+)}{U^2n_2^2+\hbar^2(\Lambda_1^--\Lambda_1^+)^2}\right.\nonumber\\
&\times\left(\rho_{n_2,2}\frac{\alpha^{n_1}}{\beta}-\rho_{n_2+1,2}\frac{\alpha^{n_1-1}}{\beta}\right) \nonumber\\   
&+\frac{2J^2(n_1+1)n_2(\Lambda_1^--\Lambda_1^+)}{U^2(n_2-1)^2+\hbar^2(\Lambda_1^--\Lambda_1^+)^2}\nonumber\\
&\left.\left(\rho_{n_2,2}\frac{\alpha^{n_1}}{\beta}-\rho_{n_2-1,2}\frac{\alpha^{n_1+1}}{\beta}\right)\right] \tag{12S}     
\end{align}   
Since  
\begin{align}
&\sum_{n_1} n_1\frac{\alpha^{n_1}}{\beta}=\frac{\Lambda_1^+}{\Lambda_1^--\Lambda_1^+} \tag{13S} \\ 
&\sum_{n_1} (n_1+1)\frac{\alpha^{n_1}}{\beta}=\frac{\Lambda_1^-}{\Lambda_1^--\Lambda_1^+} \tag{14S}
\end{align}
we obtain  
\begin{align} 
\frac{d}{dt}\rho_{n_2}=& -\frac{2J^2(n_2+1)\Lambda_1^+}{U^2n_2^2+\hbar^2(\Lambda_1^--\Lambda_1^+)^2} \rho_{n_2} \nonumber \\ 
&-\frac{2J^2n_2\Lambda_1^-}{U^2(n_2-1)^2+\hbar^2(\Lambda_1^--\Lambda_1^+)^2}\rho_{n_2} \nonumber \\ 
&+\frac{2J^2(n_2+1)\Lambda_1^+}{U^2n_2^2+\hbar^2(\Lambda_1^--\Lambda_1^+)^2}\rho_{n_2-1}\nonumber \\ 
&+\frac{2J^2n_2\Lambda_1^-}{U^2(n_2-1)^2+\hbar^2(\Lambda_1^--\Lambda_1^+)^2}\rho_{n_2+1} \tag{15S}       
\end{align}   
which is Eq.(5) of the main paper. In an analogous manner Eq.(6) can be derived.   

\subsection{Steady state for 3 sites}

Solving Eq.(6) of the main paper for the steady state (which satisfies $d {\rho_{n,2}}^{ss}/{d t} = 0$) we get a set of coupled equations
\begin{align}
 &-\sum_{l=1,3}\chi^{l,+}_{1,0} {\rho_{0,2}}^{ss}   + \sum_{l=1,3}\chi^{l,-}_{1,0} \; \rho_{1,2}^{ss} = 0 \tag{16S} \\ 
\nonumber  & -\left(\sum_{l=1,3}\chi^{l,-}_{1,0}+\chi^{l,+}_{2,1}\right) \rho_{1,2}^{ss}  +  \sum_{l=1,3}\chi^{l,-}_{2,1} \; \rho_{2,2}^{ss} \\ 
  & +  \sum_{l=1,3}\chi^{l,+}_{1,0}  \; \rho_{0,2}^{ss} = 0  \tag{17S} \\ \nonumber
          &  \vdots \\ 
 & -\left(\sum_{l=1,3}\chi^{l,-}_{n,n-1}+\chi^{l,+}_{n+1,n}\right) \rho_{n,2}^{ss} \tag{18S} \\
 \nonumber  &   +  \sum_{l=1,3}\chi^{l,-}_{n+1,n} \; \rho_{n+1,2}^{ss} +  \sum_{l=1,3}\chi^{l,+}_{n,n-1}  \; \rho_{n-1,2}^{ss} = 0 \\ \tag{19S}  
\end{align} 
Solving the first equation allows to compute 
\begin{align}
\rho_{1,2}^{ss} = \frac{\sum_{l=1,3}\chi_{1,0}^{l,+}}{\sum_{l=1,3}\chi_{1,0}^{l,-}} \;\rho_{0,2}^{ss} \tag{20S}   
\end{align}
which substituted in Eq.(17S), gives 
\begin{align}
\rho_{2,2}^{ss} =& \frac{\sum_{l=1,3}\chi_{2,1}^{l,+}}{\sum_{l=1,3}\chi_{2,1}^{l,-}}\; \rho_{1,2}^{ss} \nonumber \\ 
=&\frac{\left(\sum_{l=1,3}\chi_{2,1}^{l,+}\right)\left(\sum_{l=1,3}\chi_{1,0}^{l,+}\right)}{\left( \sum_{l=1,3}\chi_{2,1}^{l,-}\right)\left(\sum_{l=1,3}\chi_{1,0}^{l,-} \right)} \rho_{0,2}^{ss}. \tag{21S}   
\end{align}
Solving in sequence the following equations results in 
\begin{align}
\rho_{n,2}^{ss} =& \frac{ \sum_{l=1,3}\chi_{n,n-1}^{l,+}}{\sum_{l=1,3}\chi_{n,n-1}^{l,-}} \; \rho_{n-1,2}^{ss} \nonumber \\  =&
\frac{\left(\sum_{l=1,3}\chi_{n,n-1}^{l,+}\right) \dots \left(\sum_{l=1,3}\chi_{1,0}^{l,+}\right)}{\left(\sum_{l=1,3}\chi_{n,n-1}^{l,-}\right) \dots \left(\sum_{l=1,3}\chi_{1,0}^{l,-}\right)}\rho_{0,2}^{ss} \tag{22S}  
\end{align}
which is Eq.(7) of the main paper.


\begin{thebibliography}{10}
\bibitem{nosteadystates} This would not be the case for a uniform and non-interacting chain with an odd number of sites in which an even site (e.g. the second site) is connected to the environment. This is due to the fact that, in this particular case, the dissipation does not couple to one mode existing only on odd sites.
\bibitem{approx} The only approximation in these simulations is due to the local cut-off of the possible number of vibrational modes. For the parameters used here we found that a local cut-off of $20$ would not affect the results.
\bibitem{implementation} This method is implemented by approximating the density matrix of the system with matrix product states and computing the Lindbadian evolution of Eq.(\ref{eq:lind}) with a Trotter decomposition keeping a bond dimension up to $D=200$ (during the evolution the system does not reach this bond). Given the average fillings studied, and the distribution of population of the various local occupation number, we keep at least $d=10$ local states on each site. Since the system is always close to a product of local thermal states we do not keep all $d^2$ basis states but only the elements belonging to the first few main diagonals of the density matrix. Discarded elements have a weight smaller than $10^{-10}$.
\bibitem{OmegaLambda} Since in this paper $\Lambda$ is used for the coefficients of the dissipation, we use $\Omega$ to indicate the subspaces which, in \cite{SciollaKollath2015}, were identified by $\Lambda$.
\end{thebibliography}
\end{document}